\documentclass[a4paper,12pt]{article}
\usepackage{latexsym,amssymb,epsfig,rotate}
\usepackage{supertabular}
\newcommand{\cas}[0]{{\rm cas}}

\newcommand{\C}[0]{{\mathbf{C}}}

\newcommand{\N}[0]{{\mathbf{N}}}
\newcommand{\R}[0]{{\mathbf{R}}}

\newcommand{\mat}[4]{\left(\begin{array}{rr}#1 & #2\\ #3 & #4 
\end{array}\right)
}

\newcommand{\proof}{\noindent\textit{Proof.} } 
\newcommand{\diag}{\mathop{\rm diag}\nolimits}

\newcommand{\nix}[1]{}

\def\ket#1{\left|#1\right>}
\def\qed{\quad{$\Box$}}

\def\IrrP(#1){\mathop{{\rm Irr}}\nolimits(\phi\,|\,#1)}
\newcommand{\onemat}[0]{\mbox{\bf 1}}

\newcommand{\DCT}[0]{\mbox{\rm DCT}}
\newcommand{\DST}[0]{\mbox{\rm DST}}

\newcommand{\ds}{\displaystyle}

\newtheorem{theorem}{Theorem}

\newtheorem{lemma}[theorem]{Lemma}

\makeatletter
\renewcommand\@makecaption[2]{%
  \vskip\abovecaptionskip
  \sbox\@tempboxa{\small \textbf{#1:} #2}%
  \ifdim \wd\@tempboxa >\hsize
    \small \textbf{#1:} #2\par
  \else
    \hbox to\hsize{\hfil\box\@tempboxa\hfil}%
  \fi
  \vskip\belowcaptionskip}
\renewcommand \thesection {\S\@arabic\c@section}
\makeatother

\begin{document}

\title{\Large \textbf{On the Irresistible Efficiency of Signal
Processing Methods in Quantum Computing}}
\author{Andreas Klappenecker\thanks{e-mail: {\protect\tt
klappi@ira.uka.de}} ${}^{1,2}$, Martin R\"otteler\thanks{e-mail: {\protect\tt
roettele@ira.uka.de}} ${}^2$\\
\small ${}^1$Department of Computer Science, Texas A\&M University,\\[-1ex] 
\small College Station, TX 77843-3112, USA\\
\small ${}^2$Institut f{\"u}r Algorithmen und Kognitive Systeme,
Universit{\"a}t Karlsruhe,\thanks{research group Quantum Computing, Professor Thomas Beth}\\[-1ex] 
\small  Am Fasanengarten 5, D-76\,128 Karlsruhe, Germany}
%\date{\today}
\maketitle
%%%%%%%%%%%%%%%%%%%%%%%%%%%%%%%%%%%%%%%%%%
\pagestyle{empty}
\thispagestyle{empty}
%%%%%%%%%%%%%%%%%%%%%%%%%%%%%%%%%%%%%%%%%%
\begin{abstract}
We show that many well-known signal transforms allow highly efficient
realizations on a quantum computer. We explain some elementary quantum
circuits and review the construction of the Quantum Fourier Transform. We
derive quantum circuits for the Discrete Cosine and Sine Transforms,
and for the Discrete Hartley transform. We show that at most $O(\log^2 N)$
elementary quantum gates are necessary to implement any of those 
transforms for input sequences of length~$N$. 
\end{abstract}
\section{Introduction}
Quantum computers have the potential to solve certain problems at much
higher speed than any classical computer. Some evidence for this
statement is given by Shor's algorithm to factor integers in
polynomial time on a quantum computer. A crucial part of Shor's
algorithm depends on the discrete Fourier transform. The time
complexity of the quantum Fourier transform is polylogarithmic in the
length of the input signal. It is natural to ask whether other
signal transforms allow for similar speed-ups. 

We briefly recall some properties of quantum circuits and construct
the quantum Fourier transform. The main part of this paper is
concerned with the construction of quantum circuits for the discrete
Cosine transforms, for the discrete Sine transforms, and for the
discrete Hartley transform.

\section{Elementary Quantum Circuits}

The quantum computation will be done in the state space of $n$
two-level quantum systems, which is given by a $2^n$-dimensional
complex vector space. The basis vectors are denoted by $\ket{x}$ where
$x$ is a binary string of length~$n$.  The basic unit of quantum
information processing is a quantum bit or shortly qubit, which
represents the state of a two-level quantum system.

A quantum gate on $n$ qubits is an element in the group of
unitary matrices ${\cal U}(2^n)$. There are two types of gates that
are considered elementary: the XOR gates (also known as controlled
NOTs) and the single qubit operations. 

The controlled NOT gate operates on two qubits. It negates the target
qubit if and only if the control qubit is 1. Suppose that $x=b_n\dots
b_1$ is a string of $n$ bits, then 
$$ 
\mbox{CNOT}_{c,t}\ket{x}=\left\{ 
\begin{array}{ll} 
\ket{y} & \mbox{if}\quad b_c=1\\
\ket{x} & \mbox{if}\quad b_c=0 
\end{array}
\right.
$$ 
where $y$ is the bitstring obtained from $x$ by negating the bit
$b_t$. 

A single qubit gate acts on a target qubit at position $t$ by a local
unitary transformation 
$$ \onemat_{2^{n-t}}\otimes U\otimes \onemat_{2^{t-1}},\qquad U\in {\cal U}(2).$$

\begin{figure}[h]
\centerline{\epsfig{file=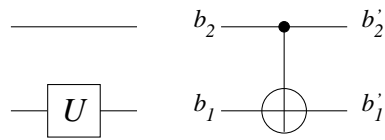,width=5cm}}
\caption{\label{elemgates} The Feynman notation for the single qubit 
gate $(\onemat_2\otimes U)$ and
for a controlled NOT operation $\ket{b_2'\,b_1'}=\ket{b_2\, b_2\oplus b_1}$.} 
\end{figure}

It will be convenient to describe the quantum circuits with a
graphical notation put forward by Feynman. The circuits are read from
left to right like a musical score. The qubits are represented by
lines, with the most significant bit at the
top. Figure~\ref{elemgates} shows the graphical notation of the elementary
gates. 

A multiply controlled NOT is defined as follows.  Let $C$ be a subset
of $[1..n]$ not containing the target $t$. Then
$$ 
\mbox{CNOT}_{C,t}\ket{x}=\left\{ 
\begin{array}{ll} \ket{y} & \mbox{if}\quad b_c=1 \mbox{ for all } c\in
C\\
\ket{x} & \mbox{otherwise} 
\end{array}
\right.
$$ 
where $\ket{y}$ is defined as above. Several controlled NOT operations
in a sequence allow us to implement the operation
$P_n\ket{x}=\ket{x+1\bmod 2^n}$. Note that $O(n)$ elementary gates are
sufficient to realize a multiply controlled NOT operation on $n$
qubits, assuming that an additional scratch qubit is
available. Therefore, at most $O(n^2)$ elementary gates are necessary
to implement the shift operation $P_n$. 
\begin{figure*}[ht]
\input{shift.pic}
\caption{\label{shift}Shift}
\end{figure*}

The state of two qubits can be exchanged with the help of three
controlled NOT operations:
$$\mbox{SWAP}_{k,h}=\mbox{CNOT}_{h,k}\mbox{CNOT}_{k,h}\mbox{CNOT}_{h,k}.$$ 
It follows that any permutation of the $n$ quantum wires can be realized
with at most $O(n)$ elementary quantum gates. 

A more detailed discussion of properties of quantum gates can be found
in~\cite{BBC+:95}. We will discuss the construction of the discrete
Fourier transform in the next section. In particular, we will show the
classical dataflow diagram and the corresponding quantum gates to
further illustrate the graphical notation.

\section{Quantum Fourier Transform}\label{sec:qft}
The discrete Fourier transform of length $N=2^n$ is defined 
by 
$$ F_N = \frac{1}{\sqrt{N}}\Big[\,\omega^{jk} \Big]_{j,k=0,\dots,N-1},$$
where $\omega=\exp(2\pi i/N)$ with $i^2=-1$. Recall the recursion step
used in the Cooley-Tukey decomposition:
\begin{eqnarray}\label{fft}
 {F_N} & = &  \Upsilon_N (\onemat_{2} \otimes F_{N/2}) 
                   \left(\begin{array}{cc} \onemat_{N/2} & \\ 
                          & T_{N/2} \end{array} \right)(F_2 \otimes
                   \onemat_{N/2})
\end{eqnarray}
where $T_{N/2}:=\diag(1,\omega, \omega^2,\dots,\omega^{N/2-1})$ denotes
the matrix of twiddle factors, and $\Upsilon_N$ denotes the permutation given
by 
$\Upsilon_N\ket{xb}=\ket{bx}$ with $x$ an $n-1$-bit integer, and $b$ a
single bit. 

We note that the implementation of $F_2$ is a local operation on a
single quantum bit. The recursion suggest four different parts of the
implementation of Fourier transforms of larger length. 
The matrix $(F_2 \otimes \onemat_{N/2})$ is a
single Hadamard operation on the most significant qubit. 
We would like to emphasize that this {\em single} quantum operation
corresponds to a full butterfly diagram. 

\begin{figure}[ht]
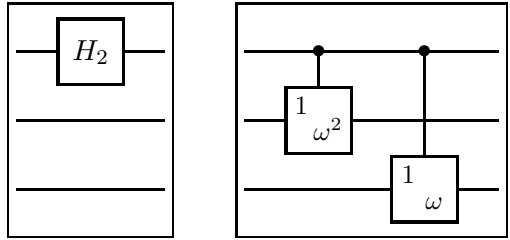

\centerline{\input{gate1.pic}\qquad\input{gate2.pic}}
\caption{\label{basicblocks} 
For length N=8, only three qubits are necessary. The circuit on the
left implements $(F_2 \otimes \onemat_{N/2})$ and the other 
realizes the twiddle matrix $\onemat_4\oplus \diag(1,\omega,\omega^2,\omega^3)$. 
}
\end{figure}

The implementation of the twiddle matrix is more complex.  
Notice that $T_{N/2}$
can be written as a tensor product of diagonal matrices
$L_j=\diag(1,\omega^{2^{j-1}})$ in the form 
$$ T_{N/2}=L_{n-1}\otimes \dots \otimes L_2\otimes L_1.$$
Thus, $\onemat_{N/2}\oplus T_{N/2}$ can be realized by 
controlled phase shift operations. Figure~\ref{basicblocks} shows the
implementation of the two operations discussed so far. 

\begin{figure}[ht]
\centerline{\input{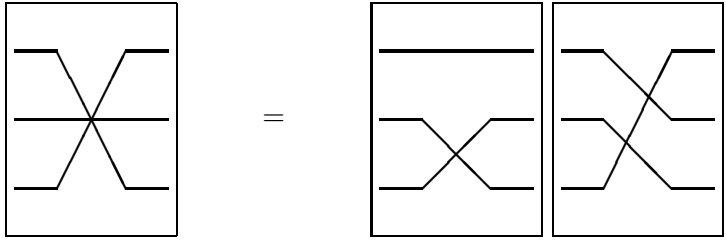}}
\caption{\label{bitreversal} The bit reversal permutation is given by 
$\onemat_2\otimes \Upsilon_4$ followed by $\Upsilon_8$.} 
\end{figure}

It remains to discuss the other two operations in (\ref{fft}).  The
operation $(\onemat_{2} \otimes F_{N/2})$ means that an implementation
of the discrete Fourier transform of length $N/2$ is used on the least
significant $(n-1)$ bits. The operation $\Upsilon_N$ is a permutation
of quantum wires. We can combine all the permutations 
$$ \Upsilon_N(\onemat_2\otimes \Upsilon_{N/2})\dots
(\onemat_{N-2}\otimes \Upsilon_4)$$ 
into a single permutation of quantum wires. The resulting permutation
is the bit reversal, see Figure~\ref{bitreversal}.
The classical and quantum implementation of the discrete Fourier
transform of length 8 are compared in Figure~5. 
We observe that the butterfly diagrams find simple realizations but
the twiddle matrices require more elementary quantum gate operations. 

\noindent The complexity of the quantum implementation can be
estimated as follows. 
If we denote by $R(N)$ the number of gates necessary to implement the
DFT of length $N=2^n$ on a quantum computer, then equation (\ref{fft})
implies the recurrence relation
$$ R(N)=R(N/2)+O(\log N)$$
which leads to the estimate $R(N)=O(\log^2N)$. 

Shor's factoring algorithm relies on the quantum Fourier transform in
a fundamental way. For more details on Fourier transforms and their
generalizations to nonabelian groups, see~\cite{Hoyer:97,PRB:99}.

\begin{figure}[p]
\rotate[l]{\epsfig{file=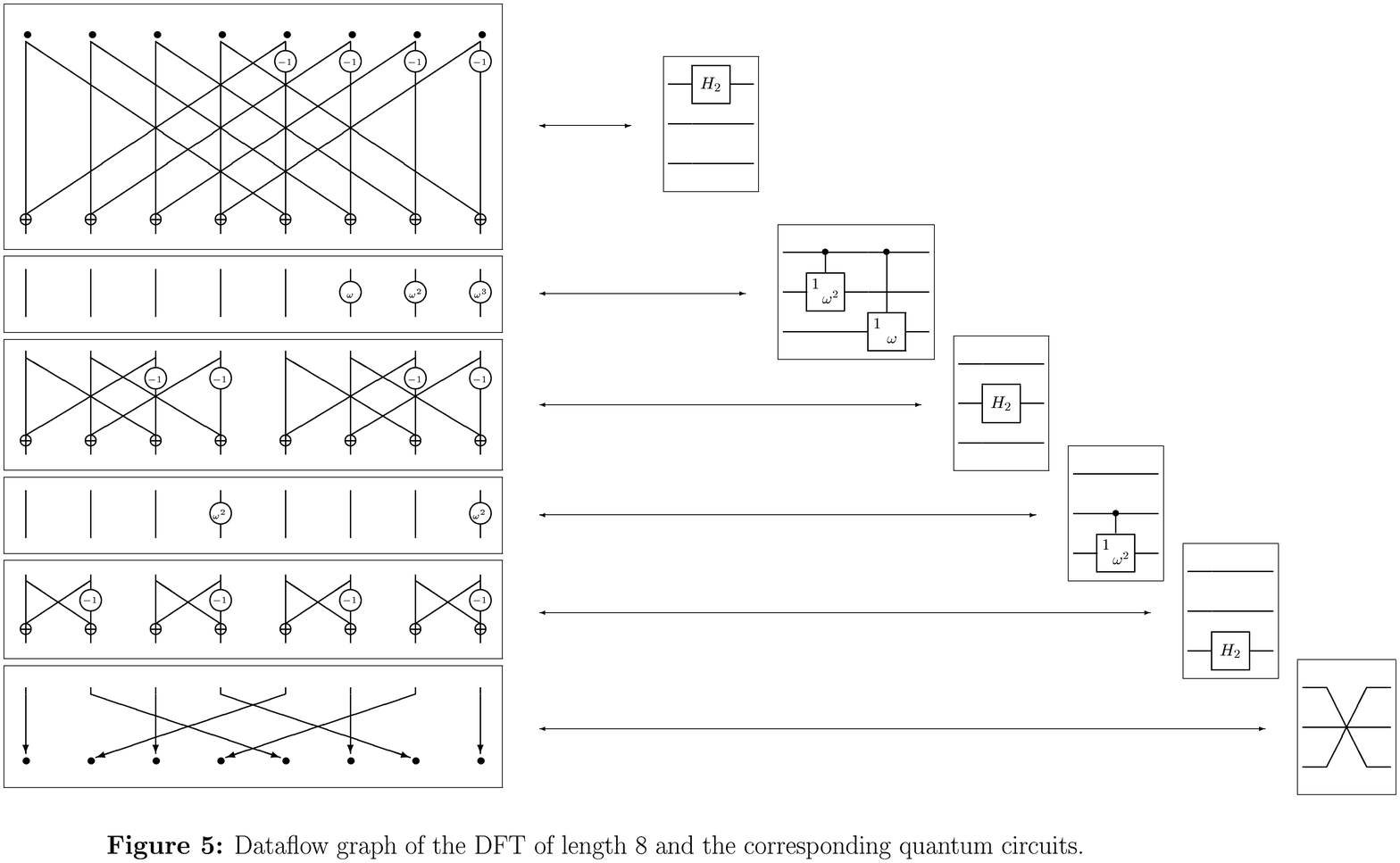,width=19.8cm}}
\end{figure}

\section{Quantum Cosine and Sine Transforms}
We derive quantum circuits for discrete Cosine and Sine transforms in
this section. The main idea is simple: reuse the circuits for the
discrete Fourier transform. 

The discrete Cosine and Sine transforms are divided into various
families. We follow~\cite{RY:90} and define the following four versions of
{\em discrete Cosine transforms}:
\[
\renewcommand{\arraystretch}{2}%
\setlength{\arraycolsep}{0.2em}%
\begin{array}{l@{\quad:=\quad}l}
C_N^{\rm I} & \left(\ds\frac{2}{N}\right)^{1/2} 
\left[k_i \cos{\ds\frac{i j\pi}{N}}\right]_
   {i, j = 0,\dots, N}
\\
C_N^{\rm II} & \left(\ds\frac{2}{N}\right)^{1/2} 
\left[k_i \cos{\ds\frac{i (j+1/2)\pi}{N}}\right]_
   {i, j = 0,\dots, N-1}
\\
C_N^{\rm III} & \left(\ds\frac{2}{N}\right)^{1/2} 
\left[k_i \cos{\ds\frac{(i+1/2) j\pi}{N}}\right]_
   {i, j = 0,\dots, N-1}
\\
C_N^{\rm IV} & \left(\ds\frac{2}{N}\right)^{1/2} 
\left[k_i \cos{\ds\frac{(i+1/2) (j+1/2)\pi}{N}}\right]_
   {i, j = 0,\dots, N-1}
\end{array}
\]
where $k_i:=1$ for $i=1,\dots,N-1$ and $k_0 := {1}/{\sqrt{2}}$.
The numbers $k_i$ ensure that the transforms are orthogonal.
The {\em discrete Sine transforms} are defined by 
\[
\renewcommand{\arraystretch}{2}%
\setlength{\arraycolsep}{0.2em}%
\begin{array}{l@{\quad:=\quad}l}
S_N^{\rm I} & \left(\ds\frac{2}{N}\right)^{1/2} 
\left[k_i \,\sin{\ds\frac{i j\pi}{N}}\right]_
   {i, j = 1,\dots, N-1}
\\
S_N^{\rm II} & \left(\ds\frac{2}{N}\right)^{1/2} 
\left[k_i \,\sin{\ds\frac{i (j+1/2)\pi}{N}}\right]_
   {i, j = 0,\dots, N-1}
\\
S_N^{\rm III} & \left(\ds\frac{2}{N}\right)^{1/2} 
\left[k_i \,\sin{\ds\frac{(i+1/2) j \pi}{N}}\right]_
   {i, j = 0,\dots, N-1}
\\
S_N^{\rm IV} & \left(\ds\frac{2}{N}\right)^{1/2} 
\left[k_i \,\sin{\ds\frac{(i +1/2)(j+1/2)\pi}{N}}\right]_
   {i, j = 0,\dots, N-1}
\end{array}
\]
where the constants $k_i$ are defined as above. Notice that 
$C_N^{\rm III}$ (resp. $S_N^{\rm III}$) is the transpose of 
$C_N^{\rm II}$ (resp. $S_N^{\rm II})$, hence it suffices to derive
circuits for the type II transforms. 

It is well-known that the trigonometric transforms can be obtained by
conjugating the discrete Fourier transform $F_{2N}$ by certain sparse
matrices. We refer the reader to 
Wickerhauser~\cite{wickerhauser93} for more details on the
decompostions.

\paragraph{DCT$_{\rm \bf I}$ and DST$_{\rm \bf I}$.}
We derive the circuits for the discrete Sine and Cosine transforms of
type I all at once. Indeed, the $\DST_{\rm I}$ and $\DCT_{\rm I}$ can be recovered from
the $\mbox{DFT}$ by a base change
\begin{equation}\label{dctI}
T_N^\dagger \cdot F_{2N} \cdot T_N = 
C_N^{\rm I} \oplus i S_N^{\rm I},
\end{equation}
where
\[
\renewcommand{\arraystretch}{1.0}%
\setlength{\arraycolsep}{0.3em}%
T_N =
\left(
\begin{array}{rrrrrrrr}
1 & & & & & & & \\
  & \frac{1}{\sqrt{2}} & & & & \frac{i}{\sqrt{2}} & &  \\
  & & \begin{picture}(5,5) 
\multiput(0,4)(2,-2){3}{\makebox(0,0){.}}
\end{picture} & & & & \begin{picture}(5,5) 
\multiput(0,4)(2,-2){3}{\makebox(0,0){.}}
\end{picture} & \\
  & & & \frac{1}{\sqrt{2}} & & & & \frac{i}{\sqrt{2}}\\
 & & & & \;1 & & & \\
  & & & \frac{1}{\sqrt{2}} & & & & {-} \frac{i}{\sqrt{2}}\\
  & &  \begin{picture}(5,5) 
\multiput(0,0)(2,2){3}{\makebox(0,0){.}}
\end{picture} & & & & \begin{picture}(5,5) 
\multiput(0,0)(2,2){3}{\makebox(0,0){.}}
\end{picture} &\\
  & \frac{1}{\sqrt{2}} & & & & {-} \frac{i}{\sqrt{2}} & &
\end{array}
\right).
\]
It is straightforward to check that (\ref{dctI}) holds, see
Theorem~3.10 in \cite{wickerhauser93}. Since we already know
efficient quantum circuits for the DFT, it remains to find an
efficient implementation of the base change matrix $T_N$.

\begin{figure*}[ht]
\input{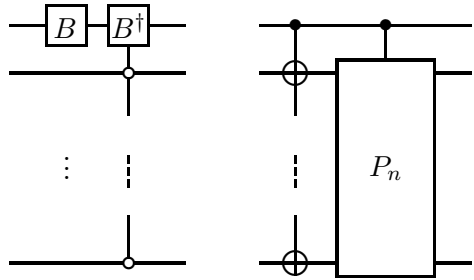}
\caption{\label{DCTIfact} Circuits realizing the
block matrix $D$ and the permutation
$\pi$.
}\label{dctIfact}
\end{figure*}

It will be convenient to denote the basis
vectors of  $\C^{2^{n+1}}$ by $\ket{b x}$, where $b$ is
a single bit and $x$ is an $n$-bit number.  
The two's complement of an $n$-bit unsigned integer 
$x$ is denoted by $x'$, that is, $x'=2^n-x$.
The action of $T_N$ can be
described by 
$$
\begin{array}{lcl@{\qquad\quad}lcl}
T_N \ket{0\mathbf{0}}&=&\ket{0\mathbf{0}} &
T_N\ket{0x}&=&\frac{1}{\sqrt{2}}\ket{0x}+\frac{1}{\sqrt{2}}\ket{1x'}\\[1ex]
T_N \ket{1\mathbf{0}}&=&\ket{1\mathbf{0}} &
T_N\ket{1x}&=&\frac{i}{\sqrt{2}}\ket{0x}-\frac{i}{\sqrt{2}}\ket{1x'}
\end{array}
$$
for all integers $x$ in the range $1\le x<2^n$. 
Ignoring the two's complement in $T_N$, we can define an operator $D$
by 
$$
\begin{array}{lcl@{\qquad\quad}lcl}
D\ket{0\mathbf{0}}&=&\ket{0\mathbf{0}} &
D\ket{0x}&=&\frac{1}{\sqrt{2}}\ket{0x}+\frac{1}{\sqrt{2}}\ket{1x}\\[1ex]
D \ket{1\mathbf{0}}&=&\ket{1\mathbf{0}} &
D\ket{1x}&=&\frac{i}{\sqrt{2}}\ket{0x}-\frac{i}{\sqrt{2}}\ket{1x}
\end{array}
$$
for all integers $x$ in the range $1\le x<2^n$. This operator is
essentially block diagonal and easy to implement by a single qubit
operation, followed by a correction. Indeed, define the matrix $B$ by $B =
\ds\frac{1}{\sqrt{2}}\mat{1}{i}{1}{-i}$, then Figure~\ref{dctIfact}
gives an implementation of the operator $D$. 

Define $\pi$ to be the permutation given by a two's complement
conditioned on the most significant bit $\pi\ket{0x}=\ket{0x}$ and
$\pi\ket{1x}=\ket{1x'}$ for all $n$-bit integers $x$.  It is clear
that $T_N=\pi D$. The circuits for the permutation $\pi$ is shown in
Figure~\ref{dctIfact}.

\begin{theorem}
The discrete Cosine transform $C_N^{\rm I}$ and the discrete Sine
transform $S_N^{\rm I}$ can be realized with at most $O(\log^2 N)$ elementary
quantum gates; the quantum circuit for these transforms 
is shown in Figure~\ref{DCTIcomplete}. 
\end{theorem}
\begin{figure*}[ht]
\input{dct1.pic}
\caption{\label{DCTIcomplete}Complete quantum circuit for the $\DCT_{\rm
I}$}
\end{figure*}
\proof Let $N=2^n$. 
We note that $O(\log^2 N)$ quantum gates are sufficient to realize the DFT of
length $2N$.  The permutation $\pi$ can be implemented with at most
$O(\log^2 N)$ elementary gates. At most 
$O(\log N)$ quantum gates are needed to realize the
operator $D$. This shows that the $\DCT_{\rm I}$ and the $\DST_{\rm I}$
can be realized with at most $O(\log^2 N)$ quantum gates. The preceding
discussion shows that Figure~\ref{DCTIcomplete} realizes the 
$\DCT_{\rm I}$ and $\DST_{\rm I}.$~\qed

\paragraph{DCT$_{\rm \bf IV}$ and DST$_{\rm\bf IV}$.} 
The trigonometric transforms of type IV are derived from the DFT by 
\begin{equation}\label{dctIV}
e^{\pi i /4N} R_N^t \cdot F_{2N} \cdot R_N = C_N^{\rm IV}
\oplus (-i)S_N^{\rm IV}.
\end{equation}
Here $R_N$ denotes the matrix 
\begin{eqnarray*}
R_N & = & \frac{1}{\sqrt{2}} \left(
\begin{array}{cccccccc}
1 & & & & -i & & & \\
  & \omega & & & & -i \omega & & \\
  & & \begin{picture}(5,5) 
\multiput(0,4)(2,-2){3}{\makebox(0,0){.}}
\end{picture} & & & & \begin{picture}(5,5) 
\multiput(0,4)(2,-2){3}{\makebox(0,0){.}}
\end{picture} & \\
& & & \omega^{N-1} & & & & -i \omega^{N-1} \\
& & & \overline{\omega}^{N} & & & & 1 \\
  & & \begin{picture}(5,5) 
\multiput(0,4)(2,2){3}{\makebox(0,0){.}}
\end{picture} & & & & \begin{picture}(5,5) 
\multiput(0,4)(2,2){3}{\makebox(0,0){.}}
\end{picture} & \\
  & \overline{\omega}^2 & & & & i \overline{\omega}^2 & & \\
\overline{\omega} & & & & i \overline{\omega}& & & \\
\end{array}
\right)
\end{eqnarray*}
with $\omega$ the primitive $4N$-th root of unity $\omega=\exp(2 \pi
i/ 4N)$. Equation~(\ref{dctIV}) is a consequence of Theorem~3.19 in
\cite{wickerhauser93} obtained by complex conjugation. 

\begin{theorem}
The discrete Cosine transform $C_N^{\rm IV}$ and the discrete Sine
transform $S_N^{\rm IV}$ can be realized with at most $O(\log^2 N)$ elementary
quantum gates; the quantum circuit for these transforms 
is shown in Figure~\ref{DCTIVcomplete}. 
\end{theorem}
\begin{figure*}[ht]
\input{dct4.pic}
\caption{\label{DCTIVcomplete}Complete quantum circuit for $\DCT_{\rm
IV}$}
\end{figure*}
\proof It remains to show that there exists an efficient quantum
circuit for the matrix $R_N$ in
equation~(\ref{dctIV}). 
A factorization of $R_N$ can be obtained as follows.  Denote by
$\overline{x}$ the one's complement of an $n$-bit integer $x$. 
We define a permutation matrix 
$\pi_1$ by $\pi_1\ket{0x}=\ket{0x}$
and $\pi_1\ket{1x}=\ket{1\overline{x}}$ for all integers $x$ in the
range of $0\le x< 2^n$. Denote by $D_1$ the diagonal matrix 
\[ 
D_1 = {\rm diag}(1, \omega, \dots, \omega^{N-1}, 
    \overline{\omega}^N, \dots, \overline{\omega}^2,
    \overline{\omega}).
\]
Then $R_N$ can be factored as 
\begin{eqnarray*}
R_N& = &
\pi_1 \cdot D_1 \cdot 
\left(\frac{1}{\sqrt{2}}
\left(
\begin{array}{rr}
 1 & -i \\
 1 &  i 
\end{array}
\right) 
\otimes \onemat_N
\right) = \pi_1\cdot D_1\cdot (\overline{B}\otimes \onemat_N).
\end{eqnarray*}

\begin{figure*}[ht]
\input{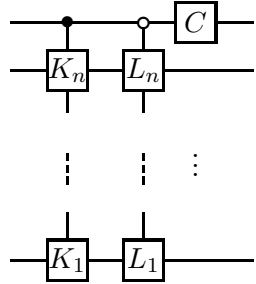}
\caption{\label{diagonal} Quantumcircuit for the diagonal matrix $D_1$. }
\end{figure*}

Note that $\overline{B}\otimes \onemat_N$ is a single qubit operation,
and $\pi_1$ can be realized by controlled not operations. The
implementation of the diagonal matrix $D_1$ is more interesting. 
Note that the diagonal matrices of increasing (decreasing) powers can
be written by tensor products 
$$ 
\begin{array}{lcl}
\Delta_1=\diag(1,\omega,\dots,\omega^{N-1})&=&L_n\otimes\cdots\otimes L_2\otimes
L_1 \\
\Delta_2=\diag(\overline{\omega}^{N-1},\dots,\overline{\omega},1)&=&K_n\otimes\cdots\otimes K_2\otimes
K_1 
\end{array}
$$
where $L_j=\diag(1,\omega^{2^{j-1}})$ and
$K_j=\diag(\overline{\omega}^{\,2^{j-1}},1)$. 
Therefore, it is possible to write $D_1$ in the form 
$D_1=(C\otimes \onemat_N)\cdot (\Delta_1\oplus\Delta_2)$ with
$C=\diag(1,\overline{\omega})$. The circuit for the diagonal matrix
$D_1$ is shown in Figure~\ref{diagonal}. 

The complete quantum
circuit for the DCT$_{\rm IV}$ is shown in
Figure~\ref{DCTIVcomplete}. Note that the last three single qubit
gates $C$, $B^\dagger$, and $M=\diag(e^{\pi i/4N},e^{\pi i/4N})$ can be
combined into a single gate $MB^\dagger C$.~\qed 

\paragraph{DCT$_{\rm \bf II}$ and DST$_{\rm \bf II}$.} The
implementation of the trigonometric transforms of type II follows a
similar pattern. Both transforms can be recovered from the DFT of
length 2N after multiplication with certain sparse matrices, 
cf.~Theorem~3.13 in~\cite{wickerhauser93}:

\begin{equation}\label{dctII}
U_N^\dagger \cdot F_{2N} \cdot V_N = 
C_N^{\rm II} \oplus (-i) S_N^{\rm II},
\end{equation}
where
\[
V_N = 
\frac{1}{\sqrt{2}}
\left(
\begin{array}{rrrrrr}
1 & & & 1 & & \\
  & \begin{picture}(5,5) 
\multiput(0,4)(2,-2){3}{\makebox(0,0){.}}
\end{picture} & & & \begin{picture}(5,5) 
\multiput(0,4)(2,-2){3}{\makebox(0,0){.}}
\end{picture} & \\
 & & 1 &  & & \phantom{-} 1 \\
 & & 1 &  & & -1\\
  & \begin{picture}(5,5) 
\multiput(0,4)(2,2){3}{\makebox(0,0){.}}
\end{picture} & & & \begin{picture}(5,5) 
\multiput(0,4)(2,2){3}{\makebox(0,0){.}}
\end{picture} & \\
1 & & & -1 &  &
\end{array}
\right)
\]
and 
\[
\renewcommand{\arraystretch}{1.2}%
\setlength{\arraycolsep}{0.3em}%
U_N =
\left(
\begin{array}{cccccccc}
1 & & & & 0 & & & \\
  & \frac{\overline{\omega}}{\sqrt{2}} & & & 
  {-} \frac{i\overline{\omega}}{\sqrt{2}} & \begin{picture}(5,5)(-20,10)
\multiput(0,4)(2,-2){3}{\makebox(0,0){.}}
\end{picture} & & \\
  & & \begin{picture}(5,5) 
\multiput(0,4)(2,-2){3}{\makebox(0,0){.}}
\end{picture} & & & \begin{picture}(5,5) 
\multiput(0,4)(2,-2){3}{\makebox(0,0){.}}
\end{picture} & &\\
  & & & \frac{\overline{\omega}^{N-1}}{\sqrt{2}} & &
  & {-} \frac{i \overline{\omega}^{N-1}}{\sqrt{2}} & 0 \\
 & & &0 &  & & & -1\\
  & & \begin{picture}(5,5)(10,10) \multiput(0,0)(2,2){3}{\makebox(0,0){.}}
\end{picture} & \frac{\omega^{N-1}}{\sqrt{2}} & & & 
  \frac{i\omega^{N-1}}{\sqrt{2}} & \\
  & &  \begin{picture}(5,5) 
\multiput(0,0)(2,2){3}{\makebox(0,0){.}}
\end{picture} & & & \begin{picture}(5,5) 
\multiput(0,0)(2,2){3}{\makebox(0,0){.}}
\end{picture} & & \\
0 & \frac{\omega}{\sqrt{2}} & & & \frac{i\omega}{\sqrt{2}} & & &
\end{array}
\right),
\]
and $\omega$ denotes the $4N$-th primitive root of unity
$\omega=\exp(2\pi i/4N)$, $i^2=-1$. 

\begin{figure*}[ht]
\input{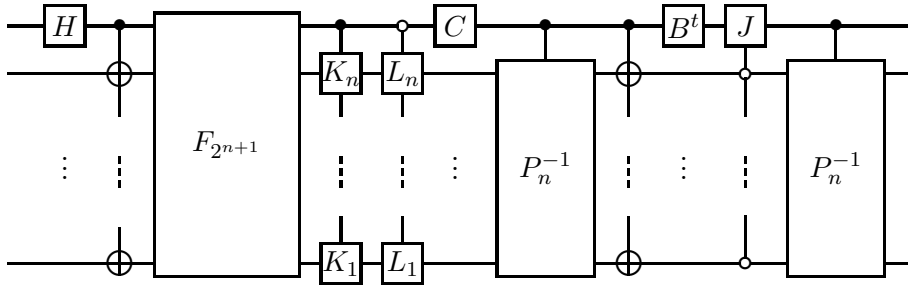}
\caption{\label{DCTIIcomplete}Complete quantum circuit for $\DCT_{\rm
II}$}
\end{figure*}

\begin{theorem}
The discrete Cosine transform $C_N^{\rm II}$ and the discrete Sine
transform $S_N^{\rm II}$ can be realized with at most $O(\log^2 N)$ elementary
quantum gates; the quantum circuit for these transforms 
is shown in Figure~\ref{DCTIIcomplete}. 
\end{theorem}
\proof We need to derive efficient quantum circuits for the matrices 
$V_N$ and $U_N$ in equation~(\ref{dctII}).
The matrix $V_N$ has a fairly simple decomposition in terms of quantum
circuits.

\begin{lemma} $V_N= \pi_1 ( H \otimes \onemat_N)$.
\end{lemma}
\proof It is clear that the Hadamard transform on the most
significant bit $H\otimes I_N$ is -- up to a permutation of rows --
equivalent to $V_N$. The appropriate permutation of rows has been introduced in the previous
section, namely $\pi_1\ket{0x}=\ket{1x}$ and $\pi_1\ket{1x}=\ket{1\overline{x}}$ for
all $0\le x< 2^n$. We can conclude that $V_N=\pi_1( H \otimes
\onemat_N)$ as desired.~\qed
\medskip

The decomposition of $U_N$ is more elaborate. Notice that
$$
\begin{array}{l@{\qquad\!}l} 
U_N\ket{0\mathbf{0}}=\ket{0\mathbf{0}} & 
U_N\ket{0 x}=\ds\frac{\overline{\omega}^{\,x}}{\sqrt{2}}\ket{0x}+
\frac{\omega^x}{\sqrt{2}}\ket{1x'}\\
U_N\ket{1\mathbf{1}}= (-1)\ket{1\mathbf{0}} &
U_N\ket{1y}=\ds -\frac{i\overline{\omega}^{\,y+1}}{\sqrt{2}}
\ket{0\,(y+1\bmod 2^n)}+\frac{i\omega^{y+1}}{\sqrt{2}}\ket{1\overline{y}} 
\end{array}
$$ 
for all integers $x$ in the range $1\le x< 2^n$ and all integers $y$
in $0\le y<2^n-1$. Here $\mathbf{0}$ and $\mathbf{1}$ denote the
$n$-bit integers $0$ and $2^n-1$ respectively. 

Define $D_0$ by $D_0\ket{1\mathbf{0}}=i\ket{1\mathbf{0}}$ and
$D_0\ket{x}=\ket{x}$ otherwise. We define a permutation $\pi_2$ by  
$\pi_2\ket{0x}=\ket{0x}$ and $\pi_2\ket{1x}=\ket{1(x+1\bmod 2^n)}$ for
all integers $x$ in $0\le x<2^n$. 
\begin{lemma}
$U_N=D_1^\dagger\, \overline{T}_N\, D_0^{\phantom{\dagger}}\, \pi_2$.
\end{lemma}
\proof
Since $D_1^\dagger\ket{0x}= \overline{\omega}^x\ket{0x}$
and  $D_1^\dagger\ket{1x}= \omega^{\,x'} \ket{1x}$, we obtain
$$ 
\begin{array}{l}
D_1^\dagger \overline{T}_N \ket{0x} = \ds \phantom{-}
\frac{\overline{\omega}^x}{\sqrt{2}}\ket{0x} + 
\frac{\omega^x}{\sqrt{2}}\ket{1x'}\\[2ex]
D_1^\dagger \overline{T}_N \ket{1x} = \ds 
-\frac{i\overline{\omega}^x}{\sqrt{2}}\ket{0x} + 
\frac{i\omega^x}{\sqrt{2}}\ket{1x'}
\end{array}
$$
We have $D_0\pi_2\ket{0x}= \ket{0x}$, $D_0\pi_2\ket{1x}=\ket{1
(x+1\bmod 2^n)}$ for all integers $x$ in $0\le x<2^n-1$, and
$D_0\pi_2\ket{1\mathbf{1}}=i\ket{1\mathbf{0}}$. We note that
$(x+1\bmod 2^n)'=\overline{x}$, whence
combining $D_1\overline{T}_N$ with
$D_0\pi_2$ shows the result.~\qed

Recall that $T_N=\pi D$. It follows that 
$$ U_N^\dagger = \pi_2^{-1} (\overline{D}_0 D^t) \pi^{-1}D_1.$$
The implementation of $D_1$ has been described in the section on the
DCT$_{\rm IV}$, and the implementation of $\pi$ (and hence $\pi^{-1}$)
is contained in the section on the DCT$_{\rm I}$. The implementation
of $\pi_2^{-1}$ is also straightforward. It remains to find an
implementation of $\overline{D}_0D^t$. We observe that
$$
\begin{array}{lcl@{\qquad\quad}lcl}
\overline{D}_0D^t\ket{0\mathbf{0}}&=&\phantom{(-i)}\ket{0\mathbf{0}} &
\overline{D}_0D^t\ket{0x}&=&\frac{1}{\sqrt{2}}\ket{0x}+\frac{i}{\sqrt{2}}\ket{1x}\\[1ex]
\overline{D}_0D^t\ket{1\mathbf{0}}&=&(-i)\ket{1\mathbf{0}} &
\overline{D}_0D^t\ket{1x}&=&\frac{1}{\sqrt{2}}\ket{0x}-\frac{i}{\sqrt{2}}\ket{1x}
\end{array}
$$\renewcommand{\arraystretch}{1.0}%
This can be accomplished by the single bit operation $B^t\otimes
\onemat_N$ followed by a multiply conditioned gate 
$J =\ds\frac{1}{\sqrt{2}}\mat{1}{-i}{-i}{1}$. 
The full circuit is shown in Figure~\ref{DCTIIcomplete}.
The statement about the complexity is clear.~\qed

\section{Quantum Hartley Transforms}\label{hartley}
The discrete Hartley transform of length $N\in \N$ is the real
$N\times N$ matrix $A_N$ defined by 
\[ A_N := \frac{1}{\sqrt{N}}\Big[ \cas\Big(\frac{2\pi i j}{N}
\Big) \Big]_{i,j=0,\ldots, N-1},
\]
where the function $\cas: \R \rightarrow \R$ is defined by
$\cas(x) := \cos(x) + \sin(x)$, see \cite{beth:89,Bracewell:79} for 
classical implementations. 
The property 
\[ A_N = \left(\frac{1-i}{2}\right) \; F_N 
    + \left(\frac{1+i}{2}\right) \; F_N^3
\] 
is easily seen from the definition. 
We derive a quantum circuit implementing $A_N$ with one
auxiliary quantum bit.

% Schaltkreis zur Hartleytransformation
% -------------------------------------
\begin{figure}[ht]
\input{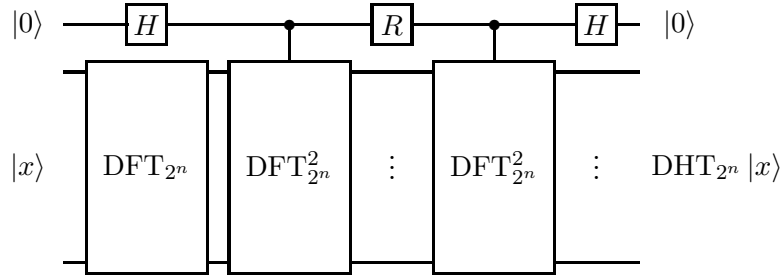}
\bigskip
\caption{\label{hartleyCirc} 
Circuit realising a quantum Hartley transform}
\end{figure}

\begin{lemma}\label{factorHart}
The discrete Hartley transform can be factorized in the form shown in
Figure \ref{hartleyCirc}. Here $R$ is the unitary circulant matrix
\[ 
R := \frac{1}{2}
\left( \begin{array}{rr} 1-i & 1+i \\ 1+i & 1-i \end{array} \right)
\]
and $H$ denotes the Hadamard transform.
\end{lemma}
\proof Let $\check{F}_N$ be the transformation which effects a DFT
conditioned to the first bit, i.\,e., written in terms of matrices we
have $\check{F}_N = \onemat_N \oplus F_N$. We now show that the given
circuit computes the linear transformation $\ket{0}\ket{x} \mapsto
\ket{0} A_N \ket{x}$ for all unit vectors $x\in \C^n$. Proceeding
from left to right in the circuit given in Figure \ref{hartleyCirc} we
obtain
\begin{eqnarray*}
\ket{0}\ket{x} 
& \stackrel{H}{\longmapsto} & \frac{1}{\sqrt{2}} (\ket{0} + \ket{1})
\ket{x} \\
& \stackrel{F_N}{\longmapsto} & \frac{1}{\sqrt{2}} (\ket{0} + \ket{1})
F_N \ket{x} \\
& \stackrel{\check{F}_N^2}{\longmapsto} & 
\frac{1}{\sqrt{2}} \ket{0} F_N \ket{x} + 
\frac{1}{\sqrt{2}} \ket{0} F_N^3 \ket{x} \\
& \stackrel{R}{\longmapsto} & 
\frac{1}{\sqrt{2}}\ket{0} \left(\frac{1}{2} (1-i) F_N + 
\frac{1}{2} (1+i) F_N^3\right) \ket{x}  \\
& \phantom{\stackrel{R}{\longmapsto}} &
+\frac{1}{\sqrt{2}}\ket{1} \left(\frac{1}{2} (1+i) F_N + \frac{1}{2} (1-i)
F_N^3 \right) \ket{x}  \\
& = & \frac{1}{\sqrt{2}} \ket{0} A_N \ket{x} +
\frac{1}{\sqrt{2}} \ket{1} F_N^{-2} A_N \ket{x}\\
& \stackrel{\check{F}_N^2}{\longmapsto} & 
\frac{1}{\sqrt{2}} (\ket{0} + \ket{1}) A_N \ket{x} \\
& \stackrel{H}{\longmapsto} & \ket{0} A_N \ket{x}
\end{eqnarray*}
as desired.~\qed

\begin{theorem}
The discrete Hartley transform $A_N$ can be computed on a
quantum computer using $O(\log^2 N)$ elementary operations if we allow one
additional ancilla qubit.
\end{theorem}
\noindent
\proof Recall that the discrete Fourier transform $F_N$
can be implemented $O(\log^2 N)$ operations as shown in Section~\ref{sec:qft}.
The statement follows from Lemma~\ref{factorHart} since all
transformations given there require at most $O(\log^2 N)$ elementary
operations.~\qed

\section{Conclusions}
We have shown that the discrete Cosine transforms, the discrete Sine
transforms, and the discrete Hartley transforms have extremely
efficient realizations on a quantum computer. All implementations
illustrated an important design principle: the reusability of highly
optimized quantum circuits. Apart from a few sparse matrices, we only
needed the circuits for the discrete Fourier transform for the
implementations. A key point is that an improvement of a basic
circuit, like the DFT, immediately leads to more efficient quantum
circuits for the DCT, DST, and DHT.

\end{document}